# CCC-predicted low-variance circles in CMB sky and LCDM

## By V. G. Gurzadyan[1] and R. Penrose[2]

1. Alikhanian National Laboratory and Yerevan State University, Yerevan, Armenia
2. Mathematical Institute, 24-29 St Giles', Oxford OX1 3LB, U.K.

*Abstract* New analysis confirms our earlier claim [1], [7] of *circles* of notably low temperature variance, often in concentric sets, in the cosmic microwave background (CMB), discernable in WMAP data. Their reality can be interpreted as evidence of supermassive black-hole encounters in a previous aeon, as predicted by conformal cyclic cosmology (CCC) [2]. Counter arguments [4-6] pointed out that such circles arise, at similar frequency, also in simulated data using WMAP's CMB power spectrum, plus random input. We responded [7] that if such circles contribute to CMB, this influences the power spectrum, enhancing such circles in simulations. We confirm this here, but show that if the *theoretical* LCDM power spectrum is used instead, then the low-variance circles disappear. This is evidence that the LCDM model gives an incomplete explanation of the CMB, missing crucial information, which is provided by incorporating low-variance circles of CCC. The excellent agreement between theoretical LCDM and observed power spectrum, even for fairly large $l$-values, does not reveal this discrepancy, of relevance only at larger $l$-values where agreement is weak. We point out various non-random aspects of the circles, seen both in the true data and in simulations with WMAP power spectrum, but not with the theoretical LCDM spectrum. We also show the spatial distribution of *concentric* circle sets to be very non-random in the true WMAP data (perhaps owing to large-scale mass inhomogeneities distorting CCC's circle shapes), in complete contrast with simulations with WMAP power spectrum, where such circle sets are much sparser and closer to average temperature. These features are fully consistent with CCC (and with an earlier analysis [8] that the random Gaussian component in the CMB is only around 0.2 in the total CMB signal) but do not readily fit in with the random initial fluctuations of standard inflation.

## 1. Introduction

In an earlier note [1] we pointed out that the Wilkinson Microwave Anisotropy Probe's 7-year data (abbreviated WMAP), and confirmed also by BOOMERanG data, contains circles which appear to have low temperature variance. Here, we provide new evidence that these circles are not merely statistical artifacts. Such circles would be in accordance with the expectations of conformal cyclic cosmology (CCC), a cosmological proposal put forward by one of us [2]. According to CCC, the current picture of the entire history of our universe, from its Big-Bang origin (but without inflation) to an indefinitely continuing exponential expansion (as would be provided by a positive cosmological constant Λ), is but one *aeon* in an unending succession of similar such aeons, where the remote future (conformal infinity $\mathscr{I}$, see [2]) of each aeon continues smoothly, as a conformal 4-manifold to become the big bang of the next. Whereas CCC does not incorporate inflation as such, the exponential expansion occurring in the previous aeon to ours would play a role that is in several respects similar to those of inflation (compare Veneziano, *et al.* [3]), but this exponential expansion occurs *prior* to the Big Bang, rather than immediately following it. The main observational distinction between CCC and the conventional inflationary picture would be that whereas in the inflationary model the initial seeds of inhomogeneity are taken to be randomly occurring quantum fluctuations, the inhomogeneities in CCC would result from a number of separate causes, in the aeon prior to ours, although all of these would be subjected to the exponential (largely self-similar) ultimate expansion of the previous aeon.

The most evident signal, particularly characteristic of CCC, would be the circles of low temperature variance referred to above, these frequently occurring in concentric sets. Such a concentric set, according to CCC, would be the result of encounters between supermassive black holes within a galactic cluster in the aeon prior to our own, each encounter resulting in an enormously intense and effectively impulsive energy burst. Such an outburst would be initially in the form of gravitational radiation, but when it reaches the conformal infinity $\mathscr{I}^\wedge$ of that previous aeon, and crosses from $\mathscr{I}^\wedge$ into the very early phase of our own aeon, its energy-momentum content is transferred, according to the equations of CCC (see [2], Appendices), into an essentially isotropic 'kick' in the initial material (taken to be the initial form of dark matter). The kick in this material, as it travels outward from the Big-Bang 3-surface $\mathscr{B}$ (which is the singularity stretched out conformally, and conformally identified with $\mathscr{I}$) until it reaches the last-scattering 3-surface $\mathscr{L}$ of our aeon, would appear to us as a circle of low variance in the CMB sky [1], [2]. See Figure 1. (The "conformal time" $t$ of Figure 1 is defined by requirements that vertical lines would describe world lines of "fundamental observers" and light rays are at 45° to the vertical, the spatial metric distance being represented uniformly along any given horizontal.) Since it is likely that a supermassive black hole in a cluster will have more than one such encounter, or that such encounters will take place for several supermassive black holes in the same cluster, we may well expect that there will be several such bursts emanating from essentially the same vertical world line (of the cluster) in the conformal diagram of Figure 1. From our own vantage point, these should appear as distinct low-variance circles that are concentric. As we shall be seeing in Section 3, there appears to be some significant evidence that the low-variance circles that are seen in the CMB data do indeed have a tendency to be clustered in concentric sets.

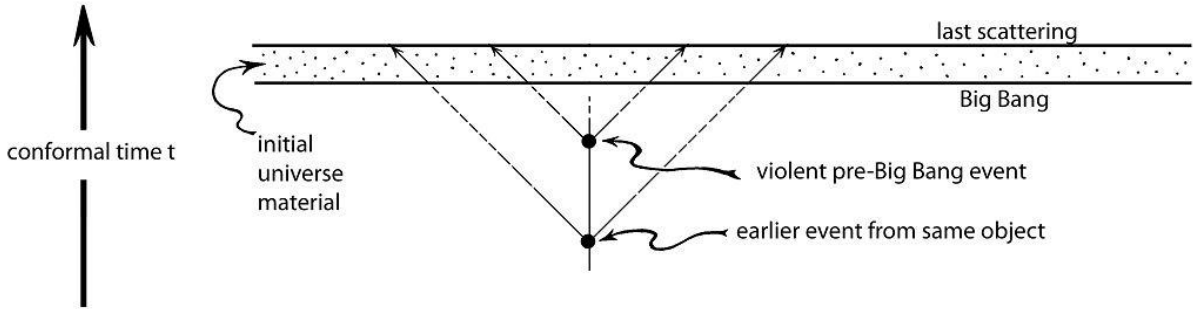

Figure 1.Conformal diagram (without inflation) according to CCC, where a pre-Big-Bang object (presumably a galactic cluster containing supermassive black-holes) is the source of two violent events.

**2. Low-variance circles in the CMB**

The search for such circles [1] was made, using WMAP by determining the temperature variance over rings (of angular width 0.5° and of angular radius in the range 2.5°–12°) for each of 10885 centres, chosen uniformly over most of the sky, but excluding a region containing the galactic disc. In [1] we used a random Gaussian model-independent calibration to demonstrate that WMAP's noise is not significant in that analysis (confirmed in [4-6]); hence the low-variance circles are indeed real features of the CMB data. The trustworthiness of this calibration was seen especially in the comparison of different noise level input in the W- and V-band and foreground reduced maps. The results were robust and confirmed that WMAP's noise has practically no role in the effect, and, accordingly, the power spectrum obtained from the WMAP data (henceforth '$l$-spectrum', where $l$ refers to the standard labeling ($l, m$) of spherical harmonics) provides a good representation of the $l$-spectrum of the actual CMB. In [4-6], a calibration was used in which this $l$-spectrum is incorporated. Since the presence of actual low-variance circles in the WMAP data would contribute to the $l$-spectrum, these circles would themselves enhance the variance of the temperature variance of the circles [4-6], so we find that incorporating the $l$-spectrum into the simulations leads to a lower sigma value (about half) that obtained using simulations that do not incorporate WMAP's $l$-spectrum. Nevertheless, as we shall see below, there are statistical features of the circles that are of more impressive significance than calibration.

It might be argued that there is a great deal more information in the temperature distribution over the whole sky than just that which is contained in the *l*-spectrum, since if we take the full spectrum of parameters (including those for *m*-values), we find that the harmonics up to a particular value *l* will be determined by $l^2$ different real numbers. Accordingly, it might seem plausible that any information carried by low-variance circles, if genuinely there in the CMB, would become lost when the *m*-value data is randomly scrambled (which is indeed part of the usual procedure) and therefore the simulations that employ the observed *l*-spectrum would not actually be biased with regard to the statistics of such circles. However, this is misleading, because there are deep theorems, due to von Neumann, Kolmogorov, and Arnold, which imply that this is not the case, the random Gaussian component being a minor fraction, around 0.2, in the CMB signal [8]. Accordingly, the *l*-spectrum does retain a memory of various statistical features in the distribution despite the scrambling of *m*-data values, so long as this is done consistently with the *l*-spectrum.

All this adds weight to the issue of whether it is appropriate to adopt the standard procedure of including the *l*-spectrum derived from the *observed* WMAP data in the estimation of the *σ*-values for the low-variance circles, in view of the circularity of this procedure, as described above. It is fortunate, therefore, that there is a straight-forward course of action for eliminating this problem, namely to use, instead, the *l*-spectrum that is predicted according to the LCDM model of standard cosmology. Since *this l*-spectrum involves no input from the low-variance circles anticipated by CCC (and uses only an initial random "seeding" for the acoustic oscillations, which may be taken, say, as originating in quantum fluctuations in an inflaton field), there should be no bias in favour of such circles in simulations which incorporate only this LCDM-predicted *l*-spectrum with randomized *m*-data. It might be felt that in view of the very close agreement between the WMAP *l*-spectrum and the LCDM-predicted *l*-spectrum—one of the great triumphs of modern cosmology—it should make little difference which *l*-spectrum is chosen. Indeed, this agreement, on the whole, is very well established [9], but there is a good deal of scope for the discrepancies between the two at large *l*-values to reflect information such as in CCC's low-variance circles, which is not part of the standard LCDM picture. Often this discrepancy is dismissed as "noise" but, as noted above, the noise element in the WMAP observations is small, and we may consider that the observed discrepancies between the WMAP data and the LCDM predictions, at high *l*-values should have genuine significance.

Indeed, we find that this is so, for we have made a comparison between three different CMB-temperature maps, with regard to the presence of low-variance circles, using

(a) the observed WMAP data;
(b) simulation using the observed WMAP *l*-spectrum and randomized *m*-data;
(c) simulation using the LCDM-calculated *l*-spectrum and randomized *m*-data.

Figure 2 displays, for circles centred at a given set of 100 randomly chosen points, with exclusion of the Galactic disk region, |b|<20° (from the 10885 referred to above), each centre colour-coded with a different colour, for radii 1°, 1.5°, 2°, 2.5°, ..., 16°, the *σ*-value (temperature standard deviation in μK) plotted vertically against the radius (in arc degrees). Figure 2a shows the plot for WMAP7 W-band data; Figure 2b is the corresponding plot for a simulation (with WMAP's noise overlapped, as usual), using the empirical WMAP *l*-spectrum and Figure 2c, using the theoretical power spectrum of the LCDM model (age of the Universe, 13.738 *Gyr*, relative densities: baryonic $\Omega_b h^2 = 0.0226$, cold dark matter $\Omega_c h^2 = 0.1140$, dark energy $\Omega_\Lambda = 0.7212$, $\Omega_{tot}=1$, Hubble parameter $H_0 = 100$ h *km s$^{-1}$ Mpc$^{-1}$*; see [9]). All plots have 1*σ* curves indicated. In Figure 3, we have picked out 10 particular randomly chosen centres from those in Figure 2, connecting them with distinctly coloured lines, to show typical behaviour of the variance for concentric sets.

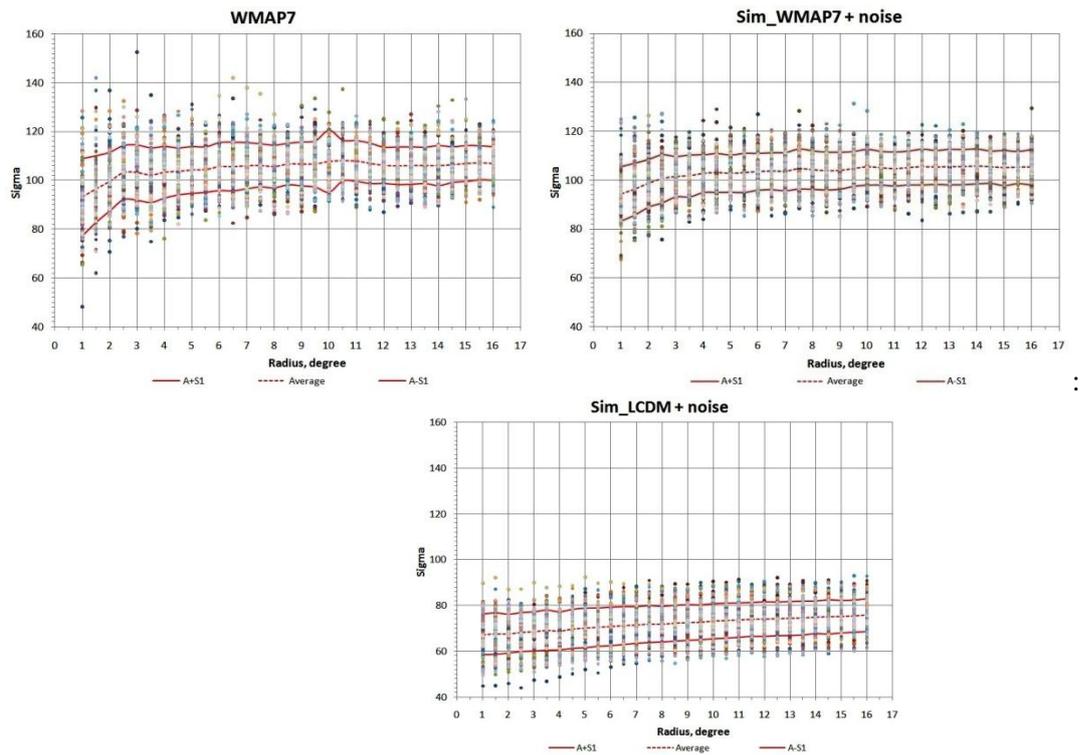

Figure 2. Temperature standard deviations vs. radius for 100 randomly chosen circle centers in (a) WMAP7 W-band map; (b) map simulated using the WMAP7 empirical power spectrum, (c) map simulated using LCDM model power spectrum; $1\sigma$ curves are indicated.

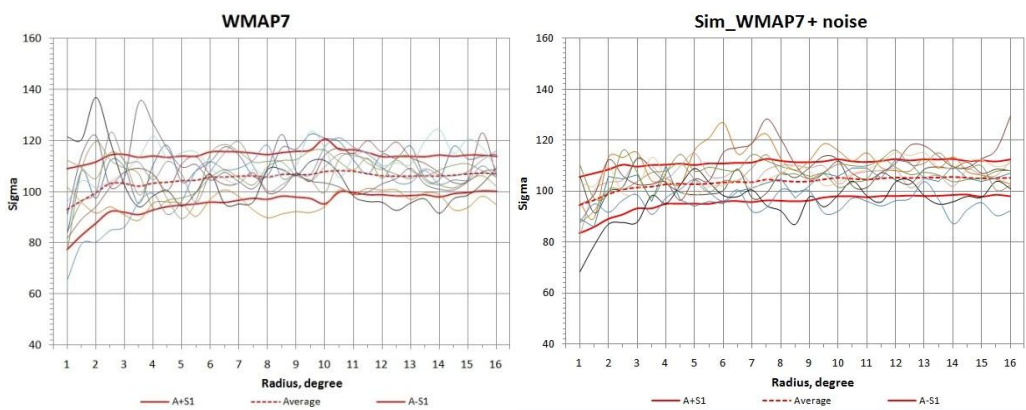

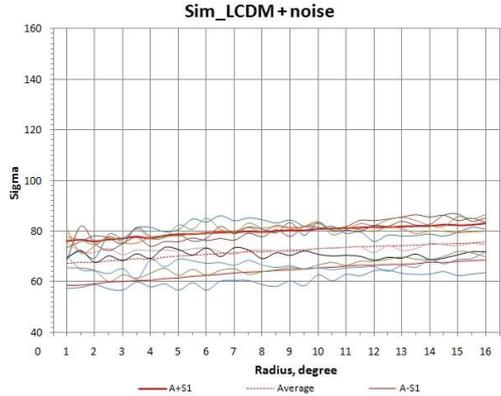

Figure 3.Ten randomly chosen curves from Fig.2 for the temperature standard deviations vs. radius; 1$\sigma$ curves are those for 100 runs as in Fig.2.

There is a striking difference between the behaviour of the low-variance circles when, on the one hand, the observed $l$-spectrum is used (Figures 3a and 3b) where low-variance dips are distinctly seen and, on the other, when the LCDM-predicted $l$-spectrum is used (Figure 3c) where low-variance circles are simply not seen. The similarity between (a) and (b) confirms the findings of our critics, that simulations incorporating the $l$-spectrum coming from the observed WMAP data (but with scrambled $m$-values) do possess low-variance circles with about the same frequency as are found in the true WMAP data (Figures 2b and 2a, respectively). However, if the theoretical $l$-spectrum of the conventional LCDM model is used instead (Figure 2c) then we find a dramatically different picture, where there is a practically linear run of the variance vs. the radius, with very little indication of low-variance circles at all.

The LCDM-based simulation (Figure 2c) shows other visible discrepancies from those (Figures 2a, 2b) using the WMAP $l$-spectrum. The most obvious is the much lower average $\sigma$-value, namely about $\sigma=70$ for the simulation using the LCDM $l$-spectrum and around $\sigma=104$ for the two plots that use the observed WMAP data. This tells us that the LCDM model gives a much "smoother" temperature distribution than either the observed data or the simulated one using the observed $l$-spectrum. This is indeed to be expected if small-scale "ripples" arise in the CMB, such as would be the case, for example, if the CCC-predicted circles are actually present in the CMB, these having various different average temperatures, and centres different from any particular one under examination, and which intersect it at various different places (as would arise in the analogous situation of ripples on a pond, following a period of rain). It may be pointed out that the CCC-predicted circles would indeed be expected to have differing average temperatures, owing to various contributing effects, most particularly a geometrical one determining to what extent the impulse (see Figure 1) is directed towards or away from us (see also Section 3).

Whether or not we accept such a picture, it is clear from the much lower average $\sigma$-value of Figure 2c that there is a lot of detail in the actual CMB that is simply not predicted by LCDM. The argument might be put forward that the higher average $\sigma$-value obtained from the use of the empirical $l$-spectrum results just from "noise" of some unspecified kind contributing to the CMB. However, this would be contradicted by the recent result that has particular relevance here, namely that the random Gaussian component is a minor fraction, around 0.2, in the CMB signal [8]. In any case, the presence of low-variance circles as seen both in the WMAP data [1] and the simulations incorporating the observed WMAP $l$-spectrum [4-6] cannot be explained merely by the addition of a random component to the calculated LCDM temperature map. For there are several indications of a distinctly non-random character in the resulting low-variance circles, as is made particularly clear from numerous features displayed in the circle distribution that are far from random, these being irrelevant to the way their significance is calibrated. We come to these next.

## 3. Depth fall-off, concentricity and the sky distribution of the circles

Whereas the actual existence of the circles claimed in [1] was confirmed in [4-6], those authors concluded that the circles are due to random fluctuations, as could have originated by the quantum fluctuations of inflation. If such an explanation were to give rise to low-variance circles, it could be expected to do so only if the resulting distribution of intensities (variance depth) of circles would have a random spatial distribution. However, we find that this is not so, and that there are various distinctive non-random features exhibited by the circles. Had the circles been of random nature, the distribution of the circles of given depth range would have been *uniform* over the radius. We see something different in the WMAP data, namely a clear fall-off, as exhibited in Figure 2a. Such a behaviour is visible also in simulations in Fig.2b and those of [4-6], but not in those of Figure 2c. Thus, this effect is determined by the correlated input information of the *l*-spectrum. Accordingly, the circle sets are of a non-random nature and cannot be random Gaussian fluctuations, as are commonly attributed to inflation. According to CCC such a correlated fall-off of the depth for chronologically earlier circles may imply relatively low energy release at initially smaller mass black hole collisions, but in any case, owing the behaviour of the conformal factor at late times in the previous aeon, such a fall-off in the intensity for the later black-hole encounters is to be expected.

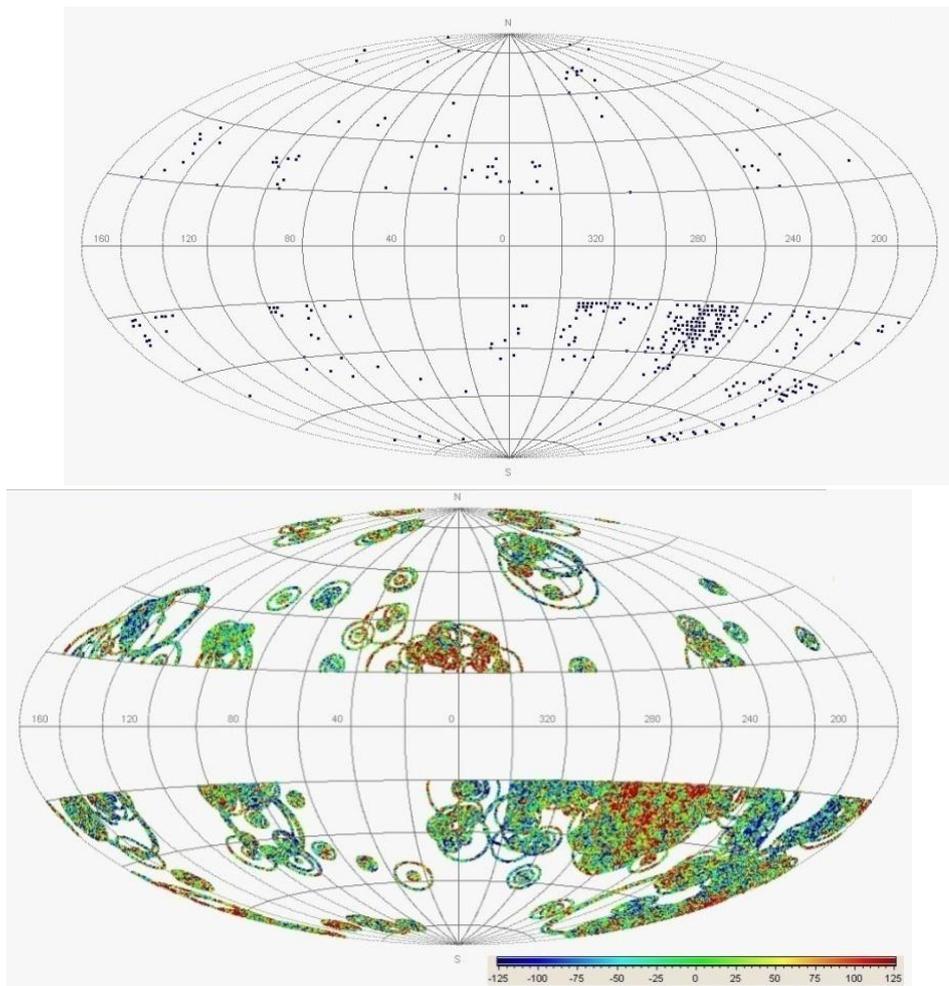

Figure 4. The sky distribution of concentric sets containing three and more circles of depth over 15μK: the upper figure indicates the positions of the centres, the lower one exhibits the actual circles.

In Figure 4, we present the distribution over the sky (without the Galactic disk region) of concentric sets containing three and more circles of over 15 μK depth; the colour bar scale is in μK within $3\sigma$ range. A distinct inhomogeneity/clustering of the sets—352 in total—is visible, which once again indicates the non-random nature of the circles. According to CCC, such a pattern could result from the interplay of two types of effect. One is that the matter distribution in the previous aeon might not be very uniform, so that the resulting distribution of (suitable) galactic clusters in the previous aeon might be far from uniform. The other (in our view, more probable) explanation is that gravitational lensing effects might be large enough that the circles resulting from black-hole encounters might become sufficiently distorted that the type of search that we are involved in here would lead to many of the energy bursts being missed. The distortions could be the result of inhomogeneities in the mass distribution in the previous aeon, or more likely, within our present aeon. Indeed, the presence of giant voids could particularly influence such images [10].

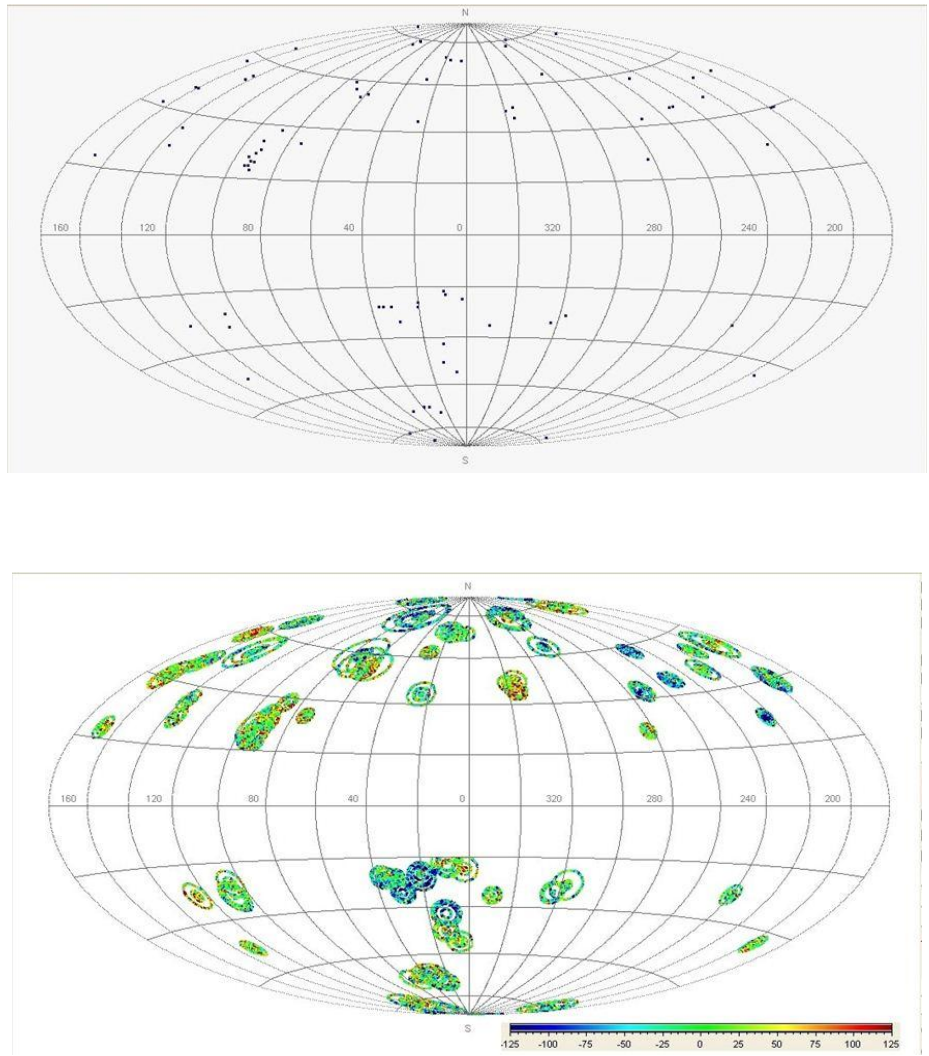

Figure 5. The corresponding maps to those of Figure 4, but where a simulated CMB sky is used incorporating WMAP's *l*-spectrum with randomized *m*-values. The differences are striking, notably the many fewer concentric sets, the absence of significant inhomogeneities and of large circles, and the much smaller departures from the average CMB temperatures.

Could LCDM together with random seeding from inflation (or from some other cause different from CCC) lead to such a picture? A spherical shape lying on the last-scattering surface $\mathscr{L}$, arising from whatever cause(not necessarily resulting from the black-hole encounters of CCC), would appear as a circle from our own vantage point if there is no subsequent lensing effect of the type just considered. Accordingly, other possible sources of the circular images, if they arise at $\mathscr{L}$, might also suffer from this kind of appearance of a non-random distribution, owing to lensing. However it is hard to see how this could have relevance to an inflationary origin for the temperature fluctuations in the CMB. It is clear from Figure 1 that inflation does not directly give rise to such disturbances, since in the conformal diagram with inflation, the source line in source line of Figure 1 would lie far into the late stages of inflation, just prior to turn-off, so the source of the disturbance would be expected to have been ironed out long before, by the inflationary expansion.

Perhaps the most compelling reason to see support for CCC in the maps presented by Figure 4 lies in the comparisons that can be made with the corresponding maps shown in Figure 5. Here, exactly the same procedure as was used to obtain the maps of Figure 4 is now applied to a *simulated* CMB sky, where WMAP's *l*-spectrum is used together with randomized *m-values*. We see several striking differences between Figures 4 and 5. Most obvious is the relative sparseness of the centres in Figure 5, where now there are only 79centres compared with the 352 of Figure 4, showing that the true data exhibits a stronger tendency for the circles to lie in concentric sets than does the simulation. Secondly, we see that there is much less crowding in certain areas, in Figure 5, giving a more random appearance, despite the far fewer centres. Thirdly, there is a distinct absence of large circles in Figure 5, compared with those in Figure 4. Finally, and perhaps most significantly, the much greener appearance of Figure 5, compared with the extremes of reds and blues in Figure 4 tells us that the average temperatures of the circles in the simulation show no particular tendency to differ much from the mean, this being in stark contrast with the situation in Figure 4, where many deviations, both warmer and cooler are seen in the true WMAP data. This is consistent, in general terms, with the expectations of CCC (see [2]), since some of the source events would have their radiation directed towards us (warmer circle-sets showing as red) and some away from us (cooler circle-sets showing as blue).

**Acknowledgements**

The authors are grateful to A.Ashtekar, and the Institute for Gravitation and the Cosmos, for financial support, to E.T.Newman, J.E.Carlstrom, and P.Ferreira for valuable discussions, and to A.L.Kashin for help with data. The use of data of WMAP, lambda.gsfc.nasa.gov, is gratefully acknowledged.